\documentclass{svproc}
\usepackage{graphicx}%
\usepackage{multirow}%
\usepackage{amsmath,amssymb,amsfonts}%
\usepackage{mathrsfs}%
\usepackage[title]{appendix}%
\usepackage{xcolor}%
\usepackage{textcomp}%
\usepackage{manyfoot}%
\usepackage{booktabs}%
\usepackage{algorithm}%
\usepackage{algorithmicx}%
\usepackage{algpseudocode}%
\usepackage{listings}%
\usepackage{ulem}
\usepackage{url}

\begin{document}
\mainmatter              
\title{On the Optimization of Methods for Establishing Well-Connected Communities}
\titlerunning{Establishing Well-Connected Communities}  
%
\author{Mohammad Dindoost\inst{1} \and 
Oliver Alvarado Rodriguez\inst{1} \and
Bartosz Bryg\inst{1} \and 
Minhyuk Park\inst{2} \and 
George Chacko\inst{2} \and 
 Tandy Warnow\inst{2} \and 
 David A. Bader \inst{1}}
\authorrunning{Dindoost \textit{et al.}} 

\institute{New Jersey Institute of Technology, Newark, NJ, USA,\\
\email{\texttt{\char`\{md724,oaa9,bb474,bader\char`\}@njit.edu}}
\and
University of Illinois Urbana-Champaign
Urbana, IL, USA,\\
\email{\texttt{\char`\{minhyuk2,chackoge,warnow\char`\}@illinois.edu}}}

\maketitle              

\begin{abstract}
Community detection plays a central role in uncovering meso scale structures in networks. However, existing methods often suffer from disconnected or weakly connected clusters, undermining interpretability and robustness. Well-Connected Clusters (WCC) and Connectivity Modifier (CM) algorithms are post-processing techniques that improve the accuracy of many clustering methods. However, they 
are computationally prohibitive on massive graphs. 
In this work, we present optimized parallel implementations of WCC and CM using the HPE Chapel programming language. 
First, we design fast and efficient parallel algorithms that leverage Chapel's parallel constructs to achieve substantial performance improvements and scalability on modern multicore architectures. Second, we integrate this software into Arkouda/Arachne, an open-source, high-performance framework for large-scale graph analytics.
Our implementations uniquely enable well-connected community detection on massive graphs with more than 2 billion edges, providing a practical solution for connectivity-preserving clustering at web scale.
For example, our implementations of WCC and CM 
enable community detection of the over 2-billion edge Open-Alex dataset in minutes using 128 cores, a result infeasible to compute previously.


\keywords{Community Detection, Complex Networks, High-Performance Computing, Parallel Algorithms} 

\end{abstract}
\section{Introduction}
Detecting community structure is a foundational problem with broad impact in science and engineering, with applications ranging from cybersecurity \cite{akoglu2015graph} to biology \cite{fortunato2010community} and social network analysis \cite{newman2004finding}. 
Numerous approaches have been developed to tackle this challenge, including graph partitioning \cite{hagen1992new,karypis1998fast}, modularity maximization and its scalable heuristics such as Louvain and Leiden \cite{newman2006modularity,blondel2008fast,traag2019louvain}, probabilistic models such as the stochastic block model (SBM) 
\cite{holland1983stochastic,abbe2018community,peixoto2019bayesian}, and flow- and motif-based techniques \cite{rosvall2008maps,benson2016higher,yang2012defining}.

Although density is often used as the main criterion for communities, ensuring they remain well connected is essential for interpretability and robustness \cite{kannan2004clusterings,traag2019louvain,park_cm_2024-journal}. However, many popular techniques can produce disconnected or weakly linked clusters: modularity-based methods can fragment groups in sparse graphs, and SBM inference may cluster weakly linked vertices together \cite{kannan2004clusterings,fortunato202220}. Such artifacts undermine interpretability and robustness. To address this, methods such as Well-Connected Clusters (WCC) \cite{park2024improved-WCC} and the Connectivity Modifier (CM)  \cite{park_cm_2024-journal,ramavarapu2024cm++} explicitly enforce user-defined intra-community connectivity standards.

WCC and CM are post-processing techniques to improve the edge-connectivity of many clustering methods, including SBMs and modularity, and have been shown to produce more accurate community structures compared to traditional approaches
\cite{park_cm_2024-journal,park2024improved-WCC}.
WCC and CM repeatedly refine and split clusters while checking connectivity, leading to significant memory and running-time costs that restrict their use to small- and medium-scale graphs but their use is computationally prohibitive for today's massive networks with billions of edges. 
High-performance frameworks such as Arkouda/Arachne \cite{rodriguez2022arachne} have demonstrated scalable solutions for other graph analytics \cite{dindoost2024vf2,dindoost2025hipermotif}, but connectivity-preserving clustering has not yet been incorporated.
We address this scalability gap by developing optimized parallel implementations of WCC and CM in the Chapel programming language \cite{chamberlain2007parallel}, integrated into the open-source Arkouda/Arachne graph analytics framework \cite{rodriguez2022arachne}. Our contributions are twofold:
\begin{enumerate}
\item \textbf{Novel Parallel Algorithms:} Chapel-based WCC and CM redesigns that reduce redundant work, increase concurrency, and achieve substantial performance improvements.
\item \textbf{Integration into Arachne:} Arachne now supports community detection with well-connectedness guarantees for practical large-scale use.
\end{enumerate}
By combining rigorous enforcement of well-connectedness with high-performance computing (HPC)-level scalability, our approach makes well-connected community detection feasible for networks at previously unattainable scale.

Our parallel implementations of WCC and CM are freely-available as open source from GitHub:  at \url{https://github.com/Bears-R-Us/arkouda-njit}.

\section{Methods for Establishing Well-Connected Communities}

In this section, we introduce highly-scalable parallel algorithms and their implementations of the WCC   and CM  methods.
The initial WCC algorithm, implemented in C++ with OpenMP parallelism, employs a shared work-queue model: the initial clusters populate a common queue accessible to all OpenMP threads, with each worker pulling the next cluster to process and pushing any new clusters generated back into the queue.
For the CM algorithm, first implemented in Python and parallelized using the multiprocessing module, the initial set of clusters is evenly divided among worker processes, each handling its assigned clusters.

Our new approach in this paper extends the original methods by incorporating scalable parallel computing strategies while preserving the  guarantees of WCC and CM that all returned clusters are well-connected. Whereas the initial implementations relied on queue-based task management, our Chapel implementations generalize this to a recursive framework that operates over a large collection of initial clusters, treating them as independent subgraphs, and thereby enabling more flexible and scalable processing.

WCC and CM share a common structure: both begin with a phase -- connected component refinement (\texttt{CCR}) -- that ensures that all input clusters are internally connected, and both apply recursive refinement to evaluate and partition clusters based on global minimum cut criteria. In both algorithms, subgraphs that satisfy a user-defined metric, typically based on cut size relative to graph size, are accepted and stored, while those that fail are recursively subdivided. The key difference lies in how they handle subgraphs that do not meet the well-connectedness criterion.
WCC always bisects such subgraphs using their minimum cut, applying further recursion to the resulting parts. 
In contrast, CM employs another approach by incorporating a user-selected community detection algorithm (referred to as CDA henceforth, where CDA can be Leiden or any other community detection method) as a refinement step. Following the established methodology, when a subgraph fails the well-connectedness criterion, after removing the min-cut to partition the cluster into two parts,  
CM applies the chosen community detection method to each resulting part to identify community structure. 
If multiple communities are found within a part, CM recurses on each community separately; otherwise, it processes the entire part as a single unit. 

This approach allows CM to identify semantically meaningful substructures by first removing weak connections and then leveraging the user's preferred community detection method to find cohesive groups within the resulting components.


As mentioned previously, both WCC and CM rely on a user-defined criterion function to determine whether a given subgraph is sufficiently well-connected. This function typically takes the form $f(n)$, where $n$ is the number of vertices in the subgraph under consideration. Common choices for $f$ include logarithmic and sublinear functions such as $\log_{10}(n)$, $\log_{2}(n)$, $\sqrt{n}$, as well as linear functions such as $kn$, where $k$ is a user-specified constant. The flexibility of this criterion allows users to tailor the sensitivity of the connectivity check to the size and structure of their input graphs.
\subsection{Optimized Parallel Implementations}
Our parallel implementations achieve performance improvements through several key optimizations tailored for large-scale graph processing in the Chapel programming language \cite{chamberlain2007parallel}. Chapel's high-level parallel constructs, including \texttt{forall} loops for parallel iteration, built-in parallel reductions, and domain-based parallelism, provide natural optimization opportunities for graph algorithms. The language's ability to handle large-scale data structures and its built-in support for parallel collections with thread-safe operations eliminate many low-level synchronization concerns while maintaining high performance.
Inside the open-source Arkouda/Arachne framework graphs are stored in double-index (DI) format \cite{rodriguez2022arachne}, which extends compressed sparse row (CSR) representation with an edge-to-source array that allows $O(1)$ edge access. This optimization is critical for our algorithms, which require frequent edge lookups.  
We employ a highly-parallel connected components algorithm to process clusters after the CCR preprocessing step. Each input cluster is evaluated in parallel, and those passing the size threshold $s_{\mbox{\textit{pre}}}$ are distributed across processing queues for concurrent evaluation. This approach leverages Chapel's parallel constructs for efficient work distribution.

To minimize overhead, we prioritize parallel cluster-level processing over parallelism within individual minimum-cut computations. This design choice recognizes that most cluster subgraphs are relatively small, making fine-grained parallelization of min-cut algorithms counterproductive. Instead, we employ a sequential variant of the VieCut \cite{henzinger2019shared} algorithm for individual subgraphs, which suffices due to their modest size, while parallelizing across the large number of clusters that require processing.

Finally, our recursive design eliminates the queue management overhead present in the original implementations, reducing memory pressure, and improving cache locality. The recursive approach naturally maps to Chapel's parallel execution model, enabling automatic work distribution and efficient memory access patterns.

\subsection{Connected Component Refinement}
The first step of both WCC and CM is to ensure that the input clusters do not contain disconnected components. Algorithm~\ref{alg:ccr} outlines the connected component refinement (\texttt{CCR}) routine, which addresses this issue.

Each cluster $c \in C$ is then converted into an induced subgraph $G_{c} = (V_{c}, E_{c})$. If $G_{c}$ contains any edges, the algorithm computes its connected components via \texttt{GetConnectedComponents}. Each connected component $cc$ with a size greater than the threshold $s_{\mbox{\textit{pre}}}$ is added to the local output queue $Q$ for further processing. This procedure ensures that only clusters composed of only one connected component are passed to the main WCC or CM routines.

\begin{algorithm}[!t]
\caption{Connected Component Refinement}
\label{alg:ccr}
\begin{algorithmic}[1]
\Procedure{\texttt{CCR}}{$G=(V,E)$, $C$, $s_{\mbox{\textit{pre}}}$}
    \State $Q \gets \varnothing$
    \ForAll{$c \in C$}
        \State $V_c \gets c$
        \State $E_c \gets \{(u,v) \in E \;:\; u \in V_c \wedge v \in V_c\}$ \Comment{Induced edges}
        \State $G_c \gets (V_c, E_c)$
        \If{$|E_c| > 0$} 
            \State $CC \gets \texttt{GetConnectedComponents}(G_c)$
            \ForAll{$cc \in CC$}
                \If{$|cc| > s_{\mbox{\textit{pre}}}$}
                    \State $Q \gets Q \cup \{cc\}$
                \EndIf
            \EndFor
        \EndIf
    \EndFor
    \State \Return $Q$
\EndProcedure
\end{algorithmic}
\end{algorithm}

\subsection{Well-Connected Clusters}

Algorithms~\ref{alg:wcc} and~\ref{alg:wcc-check} define the WCC procedure, which recursively evaluates whether clusters are internally well-connected using global minimum cuts and a user-defined criterion.
The top-level routine (Algorithm~\ref{alg:wcc}) takes a graph $G=(V,E)$, a set of clusters $C$, and size thresholds $s_{\mbox{\textit{pre}}}$ and $s_{\mbox{\textit{post}}}$. After applying the connected component refinement (\texttt{CCR}) to ensure that input clusters are connected, each resulting component $q_i$ is converted into a subgraph $G_{q_i}$ and passed to the recursive \texttt{isWCC} procedure.

The \texttt{isWCC} check (Algorithm~\ref{alg:wcc-check}) computes a global minimum cut and compares it to the threshold returned by \texttt{ComputeCriterion} (e.g., $\log_{10}(n)$ for $n=|V_c|$). If the cut size exceeds the threshold, the cluster is accepted. Otherwise, the subgraph is partitioned along the cut into $G_{c_1}$ and $G_{c_2}$, which are recursively processed if larger than $s_{\mbox{\textit{post}}}$. This hierarchical bisection continues until all accepted clusters satisfy the well-connectedness criterion. Unlike other clustering methods, no merging is performed, only recursive refinement of the input set $C$.


\begin{algorithm}[!t]
\caption{Well-Connected Clusters}
\label{alg:wcc}
\begin{algorithmic}[1]
\Procedure{\texttt{WCC}}{$G=(V,E)$, $C$, $s_{\mbox{\textit{pre}}}$, $s_{\mbox{\textit{post}}}$}
    \State $Q \gets \texttt{CCR}(G, C, s_{\mbox{\textit{pre}}})$
    \State $\mathcal{W} \gets \varnothing$ \Comment{accepted well-connected clusters}
    \ForAll{$q_i \in Q$}
        \State $V_{q_i} \gets q_i$
        \State $E_{q_i} \gets \{(u,v) \in E \;:\; u \in V_{q_i} \wedge v \in V_{q_i}\}$ \Comment{induced edges}
        \State $G_{q_i} \gets (V_{q_i}, E_{q_i})$
        \If{\Call{\texttt{isWCC}}{$G_{q_i},\, s_{\mbox{\textit{post}}}$}}
            \State $\mathcal{W} \gets \mathcal{W} \cup \{V_{q_i}\}$
        \EndIf
    \EndFor
    \State \Return $\mathcal{W}$
\EndProcedure
\end{algorithmic}
\end{algorithm}

\begin{algorithm}[!t]
\caption{The Well-Connectedness Check}
\label{alg:wcc-check}
\begin{algorithmic}[1]
    \Procedure{\texttt{isWCC}}{$G_c = (V_c, E_c)$, $s_{\mbox{\textit{post}}}$}
        \If{$|E_c| \geq 1$}
            \State $cut \gets$ \texttt{GetMinCut}($G_c$)
            \State $criterion \gets$ \texttt{ComputeCriterion}($G_c$)
            \If{$cut > criterion$}
                \State Save $V_{c}$ with a unique cluster identifier
            \Else
                \State $(V_{c_1}, V_{c_2}) \gets$ \texttt{MinCutPartition}($V_c$, $cut$)
                \State $E_{c_1} \gets \{(u,v) \in E_c : u,v \in V_{c_1}\}$
                \State $E_{c_2} \gets \{(u,v) \in E_c : u,v \in V_{c_2}\}$
                \If{$|V_{c_1}| > s_{\mbox{\textit{post}}}$}
                    \State \Call{\texttt{isWCC}}{$G_{c_1} = (V_{c_1}, E_{c_1}), s_{\mbox{\textit{post}}}$}
                \EndIf
                \If{$|V_{c_2}| > s_{\mbox{\textit{post}}}$}
                    \State \Call{\texttt{isWCC}}{$G_{c_2} = (V_{c_2}, E_{c_2}), s_{\mbox{\textit{post}}}$}
                \EndIf
            \EndIf
        \EndIf
    \EndProcedure
\end{algorithmic}
\end{algorithm}

\subsection{Connectivity Modifier}
Algorithms~\ref{alg:cm} and~\ref{alg:cm-check} define the CM procedure, which refines clusters using global min-cut and user-selected community detection algorithm (CDA), such as Leiden. As in WCC, the \texttt{CM} algorithm first applies connected component refinement (\texttt{CCR}) to produce connected subgraphs $G_{q_i}$, which are then passed to the recursive \texttt{CMC} routine.

The \texttt{CMC} check computes the global minimum cut and compares it against a user-defined threshold. If the cut exceeds the criterion, the subgraph is accepted. Otherwise, it is partitioned along the cut into $G_{c_1}$ and $G_{c_2}$. Then each part is processed with \texttt{GetCommunities}, based on user-selected Community Detection Algorithm($\rm{CDA}$).  
If multiple communities are found, they are recursively refined (subject to the $s_{\mbox{\textit{post}}}$ threshold); if not, the entire part is processed as a single unit. 
By applying community detection only after bottleneck removal, CM identifies cohesive substructures. As with WCC, refinement proceeds strictly through recursive subdivision, never merging clusters.

\begin{algorithm}[!t]
\caption{Connectivity Modifier}
\label{alg:cm}
\begin{algorithmic}[1]
\Procedure{\texttt{CM}}{$G=(V,E)$, $C$, $s_{\mbox{\textit{pre}}}$, $s_{\mbox{\textit{post}}}$, $\rm{CDA}$}
    \State $Q \gets \texttt{CCR}(G, C, s_{\mbox{\textit{pre}}})$
    \ForAll{$q_i \in Q$}
        \State $V_{q_i} \gets q_i$
        \State $E_{q_i} \gets \{(u,v) \in E \;:\; u \in V_{q_i} \wedge v \in V_{q_i}\}$ 
        \State \Call{\texttt{CMC}}{$G_{q_i} = (V_{q_i}, E_{q_i})$, $s_{\mbox{\textit{post}}}$, $\rm{CDA}$}
    \EndFor
\EndProcedure
\end{algorithmic}
\end{algorithm}

\begin{algorithm}[!t]
\caption{The Connectivity Modifier Check}
\label{alg:cm-check}
\begin{algorithmic}[1]
    \Procedure{\texttt{CMC}}{$G_c = (V_c, E_c)$, $s_{\mbox{\textit{post}}}$, $\rm{CDA}$}
        \If{$|E_c| \geq 1$}
            \State $cut \gets$ \texttt{GetMinCut}($G_c$)
            \State $criterion \gets$ \texttt{ComputeCriterion}($G_c$)
            \If{$cut > criterion$}
                \State Save $V_{c}$ with a unique cluster identifier
            \Else
                \State $(V_{c_1}, V_{c_2}) \gets$ \texttt{MinCutPartition}($V_c$, $cut$)
                \State $E_{c_1} \gets \{(u,v) \in E_c : u,v \in V_{c_1}\}$
                \State $E_{c_2} \gets \{(u,v) \in E_c : u,v \in V_{c_2}\}$
                
                \Comment{Process for each min-cut part}
                \ForAll{$G_{part} \in \{G_{c_1}, G_{c_2}\}$ where $G_{part} = (V_{part}, E_{part})$}
                    \If{$|V_{part}| > s_{\mbox{\textit{post}}}$}
                        \State $C \gets$ \texttt{GetCommunities}($G_{part}$, $\rm{CDA}$)
                        \If{$|C| > 1$} 
                            \Comment{Multiple communities found}
                            \ForAll{$V_i \in C$}
                                \State $E_i \gets \{(u,v) \in E_{part} : u,v \in V_i\}$
                                \If{$|V_i| > s_{\mbox{\textit{post}}}$}
                                    \State \Call{\texttt{CMC}}{$G_i = (V_i, E_i), s_{\mbox{\textit{post}}}$}
                                \EndIf
                            \EndFor
                        
                        \EndIf
                    \EndIf
                \EndFor
            \EndIf
        \EndIf
    \EndProcedure
\end{algorithmic}
\end{algorithm}

\section{Experimental Evaluation}
\label{sec:experiments}

To evaluate the performance and efficacy of our optimized parallel implementations of the WCC and CM algorithms, we performed a series of experiments on real-world networks. The types of experiments include: (1) performance benchmarks to assess runtime and speedup compared to the original implementations; and (2) scalability tests, encompassing strong scaling (fixed graph size, varying processors).

All experiments were performed on dual 2.0GHz AMD EPYC 7713 processors (128 cores total) with 512GB RAM. The parallel implementations were executed using Arachne \cite{rodriguez2022arachne}, while baseline comparisons utilized the original implementations.  For all experiments, we used $\log_{10}(n)$ as a user-defined criterion function, and running times were measured in wall-clock seconds.
The datasets consisted of real-world networks selected for diversity in size and structure, as detailed in Table \ref{tab:datasets}. The graphs were pre-processed, and the isolated vertices were removed. We set pre- and post-size thresholds to $s_{\textit{pre}} = s_{\textit{post}} = 1$.
We selected Leiden-CPM as the community detection algorithm (CDA) used recursively within the CM pipeline to identify communities after min-cut partitioning.


\begin{table}[htbp]
    \centering
    \caption{Real-World Networks Used in Experiments}
    \label{tab:datasets}
    \begin{tabular}{|c|c|c||c|c|c|}
        \hline
        \multicolumn{3}{|c||}{Small to Medium Networks} & \multicolumn{3}{c|}{Large Networks} \\
        \hline
        Network & vertices & Edges & Network & vertices & Edges \\
        \hline
        Bitcoin \cite{Peixoto2020Netzschleuder} & 6,336,770 & 16,057,711 & Open-Alex \cite{illinoisdatabankIDB-7362697} & 256,997,006 &  2,148,871,058 \\
        Livejournal \cite{Peixoto2020Netzschleuder} & 4,847,571 & 68,993,773 & Open-Citations-v2 \cite{illinoisdatabankIDB-5216575} & 121,052,490 & 1,962,840,983 \\
        Cit-Patents \cite{illinoisdatabankIDB-6271968} & 3,774,768 & 16,518,947 & Open-Citations \cite{illinoisdatabankIDB-0908742} & 75,025,194 & 1,363,605,603 \\
        Orkut \cite{illinoisdatabankIDB-6271968} & 3,072,441 & 117,185,083 & CEN \cite{illinoisdatabankIDB-6271968} & 13,989,436 & 92,051,051 \\
        Hyves \cite{Peixoto2020Netzschleuder} & 1,402,673 & 2,777,419 & Wikipedia-Links \cite{kunegis2013konect} & 13,593,032 & 437,217,424 \\
        \hline
    \end{tabular}
\end{table}

\subsection{Performance Benchmarks}
\label{subsec:performance}
We first compared the running time of the new WCC-Chapel and CM-Chapel implementations against the earlier WCC and CM baselines to demonstrate the efficiency gains from the novel Chapel parallelization. Experiments were conducted on networks with input clusters generated using the Leiden algorithm with Constant Potts Model using resolution parameter values of 0.001 and 0.01, reflecting two distinct modes of cluster initialization to assess performance across varying granularity levels. 

\begin{table}[htbp]
    \centering
    \caption{Runtime Comparison (in seconds) on Small to Medium Datasets}
    \label{tab:runtime_small}
    \begin{tabular}{|c|c|c|c|c|}
        \hline
        \multicolumn{5}{|c|}{Leiden CPM 0.001} \\
        \hline
        Dataset & WCC-baseline & WCC-Chapel & CM-baseline & CM-Chapel \\
        \hline
        Bitcoin & 805.9& 111.5& 65.8& 78.8\\
        Livejournal & 916.2& 87.6& 58.4& 56.1\\
        Cit-Patents & 372.3& 59.3& 43.6& 37.8\\
        Orkut & 314.0& 82.2& 88.8& 67.5\\
        Hyves & 74.8& 78.1& 18.7& 16.8\\
        \hline
        \multicolumn{5}{|c|}{Leiden CPM 0.01} \\
        \hline
        Dataset & WCC-baseline & WCC-Chapel & CM-baseline & CM-Chapel \\
        \hline
        Bitcoin & 277.1& 128.5& 116.8& 100.9\\
        Livejournal & 271.1& 120.7& 84.0& 81.1\\
        Cit-Patents & 223.1& 95.4& 66.7& 62.8\\
        Orkut & 211.9& 95.5& 74.4& 56.3\\
        Hyves & 28.3& 84.1& 26.0& 21.4\\
        \hline
    \end{tabular}
\end{table}

\begin{table}[htbp]
    \centering
    \caption{Runtime Comparison (in seconds) on Large Networks. All the dashes mean that the analysis failed due to OOM or Segmentation Faults.}
    \label{tab:runtime_large}
    \begin{tabular}{|c|c|c|c|c|}
        \hline
        \multicolumn{5}{|c|}{Leiden CPM 0.001} \\
        \hline
        Dataset & WCC-baseline & WCC-Chapel & CM-baseline & CM-Chapel \\
        \hline
        Open-Alex & - & 1306.4& - & 1317.3\\
        Open-Citations-v2 & -& 1343.9& - & 1346.2\\
        Open-Citations & - & 1230.9& - & 971.6\\
        CEN & 4330.0& 307.4& 152.3& 196.2\\
        Wikipedia-Links & - & -& 238.7& 298.9\\
        \hline
        \multicolumn{5}{|c|}{Leiden CPM 0.01} \\
        \hline
        Dataset & WCC-baseline & WCC-Chapel & CM-baseline & CM-Chapel \\
        \hline
        Open-Alex & - & 2144.7& - & 2133.73\\
        Open-Citations-v2 & -& 1891.6& - & 1850.4\\
        Open-Citations & - & 1494.7& - & 1518.3\\
        CEN & 1369.3& 240.9& 206.9& 230.0\\
        Wikipedia-Links & - & 377.2& 304.1& 372.8\\
        \hline
    \end{tabular}
\end{table}

As shown in Tables~\ref{tab:runtime_small} and \ref{tab:runtime_large}, our Chapel-based implementations demonstrate strong performance advantages on both small-to-medium- and large-scale networks.
In small-to-medium-sized networks, the new implementations show consistent and substantial improvements for WCC, achieving substantial speedups, with some networks reaching over 10x improvement. The CM results show more varied performance, with Chapel-based achieving competitive or better running time in most cases, though occasional instances favor the baseline implementation, particularly on Bitcoin where the baseline outperforms the CM-Chapel version.

The performance advantage becomes even more pronounced on large-scale networks. On massive networks such as Open-Alex and Open-Citations with more than a billion edges, the baseline implementations consistently fail due to memory limitations, whereas our Chapel-based implementations complete successfully. When both implementations can handle the dataset, as with CEN, the improvement is dramatic: WCC-Chapel achieves up to 14x speedup, while baseline CM outperforms CM-Chapel in this particular case.

These results demonstrate that our optimizations provide substantial performance improvements across the full spectrum of graph sizes, with the added critical advantage of robust scalability to networks where existing methods fail entirely. The consistent WCC performance gains and competitive CM results, combined with the ability to process billion-edge networks, establish our Chapel-based implementations as both more efficient and more capable than existing approaches. The remaining performance variations on smaller CM instances likely reflect the overhead of external C library integration for community detection (Leiden \cite{traag2019louvain}) and min-cut computation (VieCut \cite{henzinger2019shared}), direct Chapel implementations of these components would eliminate such foreign function call costs and provide even greater performance consistency.



\subsection{Scalability Analysis}
\label{subsec:scalability}
\begin{figure}[t]
    \centering
    \includegraphics[width=0.8\linewidth]{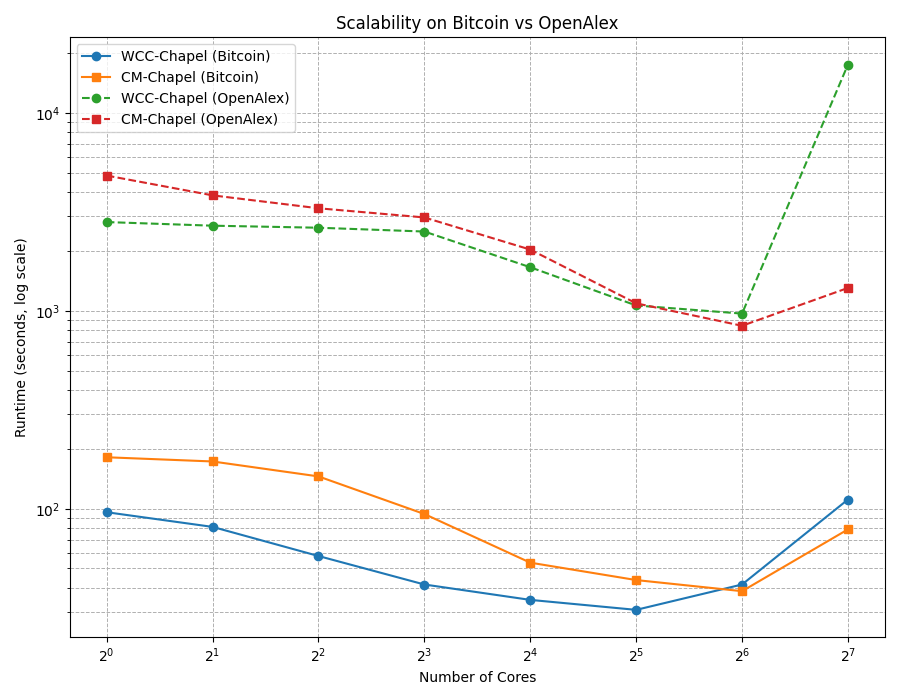}
    \caption{Strong scaling of WCC-Chapel and CM-Chapel on the Bitcoin and OpenAlex networks.}
    \label{fig:scalability}
\end{figure}
We evaluate the scalability of our implementations 
using strong scaling experiments on two representative networks: Bitcoin and Open-Alex. 
As shown in Fig.~\ref{fig:scalability}, both algorithms demonstrate clear scaling benefits up to moderate core counts before encountering performance degradation due to parallel overheads.
On the Bitcoin dataset, WCC-Chapel exhibits strong scaling from single-core up to 32 cores, achieving approximately 3x speedup at the optimal point. Beyond 32 cores, performance degrades as parallelization overheads begin to dominate the diminishing per-core workload. CM-Chapel shows similar scaling behavior but sustains improvement to 64 cores, achieving nearly 5x speedup before experiencing degradation at higher core counts. This difference suggests that the algorithmic structure of CM provides better load distribution characteristics on this particular network.

The larger Open-Alex dataset reveals more pronounced scaling differences between the algorithms. Both implementations benefit from the increased computation-to-communication ratio inherent in larger graphs, sustaining parallel efficiency to higher core counts. CM-Chapel demonstrates particularly strong scaling, achieving nearly 6x speedup at 64 cores and maintaining reasonable performance characteristics even at higher parallelization levels. 
However, WCC-Chapel exhibits a critical failure mode at maximum core count on the large network (OpenAlex), with runtime becoming dramatically worse than single-core performance. This catastrophic degradation suggests fundamental load-balancing issues or resource contention that emerge only under extreme parallelization on massive graphs. In contrast, CM-Chapel shows more graceful degradation, maintaining stability across the full range of tested core counts. These results highlight several key insights: optimal performance occurs at moderate core counts (typically 32-64), larger networks generally support higher degrees of parallelization, and algorithmic differences between WCC and CM lead to distinct scalability profiles. The severe collapse of WCC performance at high core counts on large graphs indicates that recursive refinement strategies require careful consideration of load balancing to avoid pathological behavior at scale.


\section{Conclusion}
\label{sec:conclusion}

In this work, we presented novel Chapel-based parallel implementations of community detection algorithms for WCC and CM, designed to operate efficiently on massive real-world networks. Our framework demonstrates that it is possible to combine strict connectivity guarantees with scalable performance by leveraging recursive refinement and Chapel’s parallel tasking model.  
Looking ahead, two directions are promising. First, extending this framework to distributed-memory environments will allow scaling beyond shared-memory, enabling analysis of truly web-scale networks. Second, continued optimization of Chapel-native kernels, including direct implementations of Leiden and VieCut, will reduce runtime overhead and further improve scalability. Our new parallel implementations of WCC and CM are freely-available in the open-source Arachne framework on GitHub at https://github.com/Bears-R-Us/arkouda-njit.




\section{Acknowledgments}
This research was funded in part by NSF grant numbers CCF-2109988, OAC-2402560, and CCF-2453324 (Bader) and OAC-2402559 (Warnow and Chacko).
\bibliographystyle{spmpsci} 
\bibliography{refs} 
\end{document}